\documentclass[11pt,superscriptaddress,aps,prd,preprint]{revtex4}
\usepackage{amssymb}
\usepackage{amsfonts}
\usepackage{amsmath}
\usepackage{graphicx}

\begin{document}

\title{Zeeman Effect in Phase Space}

\author{R. A. S. Paiva}\email{rendisley@gmail.com}
\affiliation{International Center of Physics, Instituto de F\'{\i}sica, Universidade de
Bras\'{\i}lia, 70910-900, Bras\'\i lia, DF, Brazil}

\author{R.G.G. Amorim} \email{ronniamorim@gmail.com}
\affiliation{International Center of Physics, Instituto de F\'{\i}sica, Universidade de
Bras\'{\i}lia, 70910-900, Bras\'\i lia, DF, Brazil}
\affiliation{Faculdade
Gama, Universidade de Bras\'{\i}lia, 72444-240, Bras\'{\i}lia, DF,
Brazil.}

\author{S. C. Ulhoa}\email{sc.ulhoa@gmail.com}
\affiliation{International Center of Physics, Instituto de F\'{\i}sica, Universidade de
Bras\'{\i}lia, 70910-900, Bras\'\i lia, DF, Brazil}

\author{A. E. Santana}\email{a.berti.santana@gmail.com}
\affiliation{International Center of Physics, Instituto de F\'{\i}sica, Universidade de
Bras\'{\i}lia, 70910-900, Bras\'\i lia, DF, Brazil}

\author{F. C. Khanna \footnote{Professor Emeritus - Physics Department, Theoretical Physics Institute, University of Alberta\\
Edmonton, Alberta, Canada}}
\email[]{khannaf@uvic.ca; fkhanna@ualberta.ca}
\affiliation{Department of Physics and Astronomy, University of
Victoria,BC V8P 5C2, Canada}

\begin{abstract}
The two-dimensional hydrogen atom in an external magnetic field is considered in the context of phase space. Using solution of the Schr\"{o}dinger equation in phase space the Wigner function related to the Zeeman effect is calculated. For this purpose, the Bohlin mapping is used to transform the Coulomb potential into a harmonic oscillator problem. Then it is possible to solve the Schr\"{o}dinger equation easier by using the perturbation theory. The negativity parameter for this system is realised. 
\end{abstract}

\keywords{Moyal product; Phase space; Wigner function, Zeeman effect}
\pacs{03.65.Ca; 03.65.Db; 11.10.Nx}

\maketitle

\section{Introduction}

Since the early years of the development of quantum mechanics its formulation in phase space has been troublesome. The seminal paper by Wigner in 1932 addressed such a problem in an attempt to deal with the superfluidity of Helium~\cite{wig1}. The Wigner function, $f_{W}(q,p)$, was introduced as a Fourier transform of the density matrix, $\rho(q,q^{\prime})$. Then the phase space manifold ($\Gamma$), which has a symplectic structure, is described by the coordinates $(q,p)$~\cite{wig1,wig2,wig3,wig4}. The Wigner function is identified as a
quasi-probability density since $f_{W}(q,p)$ is real but it is not
positive defined. However, the integrals $\rho(q)=\int f_{W}(q,p)\,dp$ and $\rho(p)=\int f_{W}(q,p)\,dq$ are distribution functions.

In the Wigner formalism, each operator, $A$, defined in the Hilbert
space, $\mathcal{H}$, is associated with a function, $a_{W}(q,p)$, in $\Gamma
$. This procedure is precisely specified by a mapping $\Omega_{W}:A\rightarrow
a_{W}(q,p)$, such that, the associative algebra of operators defined in
$\mathcal{H}$ leads to be an algebra in $\Gamma,$ given by $\Omega
_{W}:AB\rightarrow a_{W}\star b_{W},$ where the star-product, $\star$,$\,$\ is
defined by
\begin{equation}
a_{W}\star b_{W}=a_{W}(q,p)\exp\left[  \frac{i\hbar}{2}(\frac{\overleftarrow
{\partial}}{\partial q}\frac{\overrightarrow{\partial}}{\partial p}%
-\frac{\overleftarrow{\partial}}{\partial p}\frac{\overrightarrow{\partial}%
}{\partial q})\right]  b_{W}(q,p). \label{mar261}%
\end{equation} 

As a consequence a non-commutative structure in
$\Gamma$ is obtained, that has been explored in different ways~\cite{wig2}-\cite{seb13}.
Recently \cite{seb1,seb2,seb22,sig1,seb222} unitary representations of
symmetry Lie groups have been obtained on a symplectic manifold, exploring
the noncommutative nature of the star-product and using
the mapping $\Omega_{W}$~\cite{seb1,seb2,seb22}. As a result, using a specific representation of a Galilei group, Schr\"{o}dinger equation in phase space is obtained. On the other hand, the scalar representation of
Lorentz group for spin 0 and spin 1/2, leads to the
Klein-Gordon and Dirac equations in phase space. In relativist and non-relativistic approach, the wave functions
are closely associated with the Wigner function~\cite{seb1,seb2}. This
provides a fundamental ingredient for the physical interpretation of the
formalism showing its advantage in relation to other attempts.

In recent years the two-dimensional physical systems have been investigated due to both experimental and theoretical interests. For example, it is possible to cite the fractional Hall effect in a tilted magnetic field \cite{h1,h2}, superconductivity in two-dimensional organic conductors induced by magnetic field \cite{h3}, investigations of graphene \cite{h4,h5}, etc. Particulary, two-dimensional model of hydrogen atom was considered in several contexts, for instance it is possible to describe highly three-dimensional anisotropic crystals \cite{h6}, semiconductor heterostructures \cite{h7,h8, h9} and astrophysical applications \cite{h10, h11, h12}. In addition, the hydrogen atom in an uniform magnetic field can present a non-classical behavior with the increase of strength of magnetic field \cite{h13, h14, h15, h16}. In this paper, the objective is to investigate the two-dimensional hydrogen atom in a constant magnetic field in phase space picture, i.e., the Schr\"{o}dinger equation in phase space is used to study the Zeeman effect. Zeeman effect is applied in a variety of systems, including intense laser lights, comic rays, and the study  of intergalactic and interstellar medium \cite{1,2}. The importance to study the Wigner function for such an effect is in order to obtain information about the chaotic nature of such systems.

In Section \ref{sec1}, a review about the formalism of quantum mechanics in phase space and its connection with Wigner function is presented. In Section \ref{sec2}, the hamiltonian of hydrogen atom in an external magnetic field and Bohlin mapping are discussed. The solution for the Schr\"{o}dinger equation for Zeeman effect in phase space is solved by perturbative method in Section \ref{sec3}. In Section \ref{sec4} a summary and concluding remarks are presented.

\section{Symplectic Quantum Mechanics}\label{sec1}

In this section, the representation of the Galilei group in $\mathcal{H}(\Gamma )$ is presented.
This procedure leads us to the Schr\"odinger equation in phase space. Then a
connection between this representation and the Wigner formalism is established.

Using the star-operator, $\widehat{A}=a\star $, the position and momentum
operators, respectively, are defined by
\begin{equation}
\widehat{Q}=q\star =q+\frac{i\hbar }{2}\partial _{p},  \label{eq 7}
\end{equation}

\begin{equation}
\widehat{P}=p\star =p-\frac{i\hbar }{2}\partial _{q}.  \label{eq 8}
\end{equation}%

The operators given in Eqs.(\ref{eq 7}-\ref{eq 8}) satisfy the Heisenberg commutation relation,
\begin{equation}\nonumber
[\widehat{Q},\widehat{P}]=i\hbar.
\end{equation}

In addition, the following operators are introduced,

\begin{equation}
\widehat{K}=m\widehat{Q}_{i}-t\widehat{P}_{i},  \label{eq 9}
\end{equation}
\begin{equation}\label{ma}
\widehat{L}_{i} =\epsilon _{ijk}\widehat{Q}_{j}\widehat{P}_{k},  
\end{equation}

\begin{equation}\label{ham}
\widehat{H} =\frac{\widehat{P}^{2}}{2m}=\frac{1}{2m}(\widehat{P}_{1}^{2}+%
\widehat{P}_{2}^{2}+\widehat{P}_{3}^{2}).  
\end{equation}%

From this set of unitary operators, after simple calculations, the following set of commutation relations are obtained:

\begin{equation*}
\lbrack \widehat{L}_{i},\widehat{L}_{j}]=i\hbar \epsilon _{ijk}\widehat{L}%
_{k},
\end{equation*}%
\begin{equation*}
\lbrack \widehat{L}_{i},\widehat{K}_{j}]=i\hbar \epsilon _{ijk}\widehat{K}%
_{k},
\end{equation*}%
\begin{equation*}
\lbrack \widehat{L}_{i},\widehat{P}_{j}]=i\hbar \epsilon _{ijk}\widehat{P}%
_{k},
\end{equation*}%
\begin{equation*}
\lbrack \widehat{K}_{i},\widehat{P}_{j}]=i\hbar m\delta _{ij}\mathbf{1},
\end{equation*}%
\begin{equation*}
\lbrack \widehat{K}_{i},\widehat{H}]=i\hbar \widehat{P}_{i},
\end{equation*}%
with all other commutation relations being null. This is the Galilei-Lie algebra with a central extension characterized by $m$. The operators defining the Galilei symmetry $\widehat{P}$, $\widehat{K}$, $\widehat{L}$ and $\widehat{H}$ are then generators of translations, boost, rotations and time translations, respectively.




Defining the operators
\begin{equation*}
\overline{Q}=q\mathbf{1}\ \ \ \mathrm{and}\ \ \overline{P}=p\mathbf{1},
\end{equation*}%
for boost, operators $\overline{Q}$ and $\overline{P}$,
transform as

\begin{equation*}
\exp (-iv\frac{\widehat{K}}{\hbar })2\overline{Q}\exp (iv\frac{\widehat{K}}{%
\hbar })=2\overline{Q}+vt\mathbf{1},
\end{equation*}

and

\begin{equation*}
\exp (-iv\frac{\widehat{K}}{\hbar })2\overline{P}\exp (iv\frac{\widehat{K}}{%
\hbar })=2\overline{P}+mv\mathbf{1}.
\end{equation*}%
This shows that, $\overline{Q}$ and $\overline{P}$ transform as position and
momentum variables, respectively. These operators satisfy $[\overline{Q},%
\overline{P}]=0$. Then $\overline{Q}$ and $\overline{P}$ cannot be
interpreted as observables. Nevertheless, they can be used to construct a Hilbert space framework in phase space. Then we define the functions $\phi(q,p)$ 
in $\mathcal{H}(\Gamma )$, that satisfy the condition
\begin{equation*}
\int dq dp \phi^{\ast}(q,p)\phi(q,p)<\infty.
\end{equation*}

The
wave function $\psi (q,p,t)=\langle q,p|\psi (t)\rangle $  associated with
the state of the system is defined, but do not have the content of the usual quantum mechanics state.

The time evolution equation for $\psi (q,p,t)$ is derived by using the
generator of time translations, i. e.
\begin{equation}
\psi (t)=e^{\frac{-i\widehat{H}t}{\hbar }}\psi (0).  \label{eq 12}
\end{equation}%

Then this leads to

\begin{equation*}
i\hbar \partial _{t}\psi (q,p;t)=\widehat{H}(q,p)\psi (q,p;t),
\end{equation*}%
or
\begin{equation}
i\hbar \partial _{t}\psi (q,p;t)=H(q,p)\star \psi (q,p;t),  \label{eq 13}
\end{equation}%
which is the Schr\"odinger equation in phase space~\cite{seb1}.

The association of $\psi (q,p,t)$ with the Wigner function is given by \cite%
{seb1},
\begin{equation}
f_{W}(q,p)=\psi (q,p,t)\star \psi ^{\dagger }(q,p,t).  \label{eq14}
\end{equation}%
This function satisfies the Liouville-von Neumann equation \cite{seb1}.

\section{two-dimensional hydrogen atom in an external magnetic field and Bohlin Mapping}\label{sec2}
 
The Hamiltonian for the two dimensional hydrogen atom in a constant and uniform magnetic field  $\mathbf{B}=B\widehat{z}$ is given as \cite{hid1, hid2, hid3}
\begin{equation}\label{b1}
H=\frac{(\mathbf{P}-e\mathbf{A})^2}{2m}-\frac{k}{(x^2+y^2)^{1/2}},
\end{equation}
where $m$ and $e$ represent the electron mass and charge, respectively, $A$ is the magnetic potential vector and $k$ is a constant. In two-dimensional case Eq.(\ref{b1}) is written as
\begin{equation}\label{b2}
H=\frac{\mathbf{P}^2}{2m}-\frac{k}{(x^2+y^2)^{1/2}}+\frac{m\omega}{2}(x^2+y^2)+\omega L_z,
\end{equation}
where $\omega=eB/2mc$ is the frequency and $L_z$ is angular momentum in the $z$-direction. Here the constant $\omega L_z$ will be neglected once the energy is defined up to a constant.

In order to solve the Schr\"{o}dinger equation for this Hamiltonian, the Bohlin mapping is used.

\subsection{Bohlin mapping}

Bohlin mapping is defined by \cite{boh1, boh2, boh3}
\begin{equation}\label{b3}
x+iy=(q_1^2-q_2^2)+i(2q_1q_2),
\end{equation}
or
\begin{equation}\label{b4}
x=q_1^2-q_2^2,
\end{equation}
and
\begin{equation}\label{b5}
y=2q_1q_2.
\end{equation}

Defining
\begin{equation}\label{b6}
P_{x}+iP_{y}=\frac{p_1+ip_2}{2(q_1+iq_2)},
\end{equation}
leads to
\begin{equation}\label{b7}
P_x=\frac{p_1q_1+p_2q_2}{2(q_1^2+q_2^2)},
\end{equation}
and
\begin{equation}\label{b8}
P_y=\frac{p_2q_1-p_1q_2}{2(q_1^2+q_2^2)}.
\end{equation}
Substituting Eqs.(\ref{b4}), (\ref{b5}), (\ref{b7}) and (\ref{b8}) in Eq.(\ref{b2}), leads to the Hamiltonian
\begin{equation}\label{b9}
H=\frac{1}{2}\frac{p_1^2+p_2^2}{(q_1^2+q_2^2)}-\frac{k}{(q_1^2+q_2^2)}+\frac{B^2}{8}(q_1^2+q_2^2)^3.
\end{equation}
Using $\hbar=\omega=e=m=1$ and taking the hyper-surface given by $H=E$, leads to

\begin{equation}\label{b10}
\frac{1}{2}(p_1^2+p_2^2)+\frac{B^2}{8}(q_1^2+q_2^2)^3-E(q_1^2+q_2^2)-k=0\,,
\end{equation}
which is the Hamiltonian to be used in the next section. It should be noted that Bohlin transformation is a canonical transformation \cite{bohlincan}.

\section{Zeeman effect in phase space}\label{sec3}

Using Eq. (\ref{b10}), the equation is written as
\begin{equation}\label{b11}
\left[\frac{1}{2}(p_1^2+p_2^2)+\frac{B^2}{8}(q_1^2+q_2^2)^3-E(q_1^2+q_2^2)-k\right]\star\psi(q_1,p_1,q_2,p_2)=0.
\end{equation}
It should be noted that the above equation is obtained from the classical Hamiltonian by means of the star product. Thus the Bohlin mapping that lead to Eq. (\ref{b10}) is a classical transformation. This equation is analyzed by perturbation theory. The equation in phase space is defined as
\begin{equation}\label{12}
(\widehat{H}_0+\widehat{H}_1)\psi(q_1,p_1,q_2,p_2)=k\psi(q_1,p_1,q_2,p_2),
\end{equation}
where $\widehat{H}_0=\frac{1}{2}(p_1^2+p_2^2)\star-E(q_1^2+q_2^2)\star$ and $\widehat{H}_1=\frac{B^2}{8}(q_1^2+q_2^2)^3\star$.

The equation for $\widehat{H}_0$ has the form
\begin{equation}\label{b13}
\widehat{H}_0\psi^{(0)}(q_1,p_1,q_2,p_2)=k^{(0)}\psi^{(0)}(q_1,p_1,q_2,p_2),
\end{equation}
 where $\psi^{(0)}(q_1,p_1,q_2,p_2)$ and $k^{(0)}$ represent, respectively, the eigenfunction and eigenvalue of the unperturbed hamiltonian. 

Defining the operators
\begin{equation}\label{b14}
\widehat{a}=\left(\sqrt{\frac{W}{2}}q_1\star+i\sqrt{\frac{1}{2W}}p_1\star\right),
\end{equation}
\begin{equation}\label{b15}
\widehat{a}^\dagger=\left(\sqrt{\frac{W}{2}}q_1\star-i\sqrt{\frac{1}{2W}}p_1\star\right),
\end{equation}
\begin{equation}\label{b16}
\widehat{b}=\left(\sqrt{\frac{W}{2}}q_2\star+i\sqrt{\frac{1}{2W}}p_2\star\right),
\end{equation}
\begin{equation}\label{b17}
\widehat{b^{\dagger}}=\left(\sqrt{\frac{W}{2}}q_2\star-i\sqrt{{1}{2W}}p_2\star\right),
\end{equation}
where $W^2/2=-E$, the star operators  $q_{i}\star$ and $p_{i}\star$ are given by
\begin{equation}\label{b18}
q_{i}\star=q_{i}+\frac{i}{2}\frac{\partial}{\partial p_i},
\end{equation}
\begin{equation}\label{b19}
p_{i}\star=p_{i}-\frac{i}{2}\frac{\partial}{\partial q_i},
\end{equation}
and perturbed hamiltonian is 
\begin{equation}\label{b20}
\widehat{H}=W(\widehat{a}\widehat{a}^{\dagger}+\widehat{b}\widehat{b}^{\dagger}+1)+\frac{B^2}{8}[(\widehat{a}+\widehat{a}^{\dagger})^2+(\widehat{b}+\widehat{b}^{\dagger})^2]^3.
\end{equation}
The unperturbed hamiltonian is defined as 
\begin{equation}\label{b21}
\widehat{H}_0=W(\widehat{a}\widehat{a^{\dagger}}+\widehat{b}\widehat{b^{\dagger}}+1).
\end{equation}
Then, the perturbed part is
\begin{equation}\label{b22}
\widehat{H}_1=\frac{B^2}{8}[(\widehat{a}+\widehat{a}^{\dagger})^2+(\widehat{b}+\widehat{b}^{\dagger})^2]^3.
\end{equation}
The equation that is to be analyzed is given as
\begin{equation}\label{b23}
H\star\psi(q_1,p_1,q_2,p_2)=k\psi(q_1,p_1,q_2,p_2).
\end{equation}
The unperturbed equation is
\begin{equation}\label{b24}
H_0\star\psi_{n_1,n_2}^{(0)}(q_1,p_1,q_2,p_2)=k^{(0)}_{n_1,n_1}\psi_{n_1,n_2}^{(0)}(q_1,p_1,q_2,p_2),
\end{equation}
The unperturbed part, $\widehat{H}_0$, has solutions given by
\begin{equation}\label{b25}
\psi_{n_1,n_2}^{(0)}(q_1,p_1,q_2,p_2)=\phi_{n_{1}}(q_1,p_1)\Gamma_{n_{2}}(q_2,p_2),
\end{equation}
where $\phi_{n_{1}}(q_1,p_1)$ and $\Gamma_{n_{2}}(q_2,p_2)$  are solutions.  The eigenvalue equations are given by 
\begin{equation}\label{b26}
\widehat{a}\phi_{n_1}=\sqrt{n_1}\phi_{n_1-1},
\end{equation}
\begin{equation}\label{b27}
\widehat{a^\dagger}\phi_{n_1}=\sqrt{n_1+1}\phi_{n_1+1},
\end{equation}
\begin{equation}\label{b28}
\widehat{b}\Gamma_{n_2}=\sqrt{n_2}\Gamma_{n_2-1},
\end{equation}
\begin{equation}\label{b29}
\widehat{b^\dagger}\Gamma_{n_2}=\sqrt{n_2+1}\Gamma_{n_2+1}.
\end{equation}
Using the relations $\widehat{a}\phi_{0}=0$ and $\widehat{b}\Gamma_{0}=0$, the ground-state solution is, 
\begin{equation}\label{b30}
\psi_{0,0}^{(0)}(q_1,p_1,q_2,p_2)=\mathcal{N}e^{-(Wq_{1}^2+p_{1}^2)}L_{n_1}(Wq_{1}^2+p_{1}^2)e^{-(Wq_{2}^2+p_{2}^2)}L_{n_2}(Wq_{2}^2+p_{2}^2),
\end{equation}
 where $L_{n_1}$ and $L_{n_2}$ are Laguerre polynomials; and  $\mathcal{N}$ is a normalization constant.  The eingenvalues solutions, given in Eq.(\ref{b24}), are
\begin{equation}\label{b31}
k^{(0)}_{n_1,n_2}=(n_1+n_2+1)W.
\end{equation}
 The excited states are obtained from Eq.(\ref{b30}) using operators given in Eqs. (\ref{b26}-\ref{b28}). 

Then, the solution for the first order perturbed Hamiltonian is given by

\begin{eqnarray}\label{b32}
\psi_{n_1,n_2}^{(1)}(q_1,p_1,q_2,p_2)&=&\psi_{n_1,n_2}^{(0)}(q_1,p_1,q_2,p_2)\nonumber\\&+&\sum_{m_1\neq n_1; m_2\neq n_2}\left(\frac{\int \psi_{m_1,m_2}^{\ast(0)}(q_1,p_1,q_2,p_2)\widehat{H}_{1}\psi_{n_1,n_2}^{(0)}(q_1,p_1,q_2,p_2)dq_1dp_1dq_2dp_2}{k_{n_1,n_2}^{(0)}-k_{m_1,m_2}^{(0)}}\right)\nonumber\\
&\times&\psi_{m_1,m_2}^{(0)}(q_1,p_1,q_2,p_2).
\end{eqnarray}
It is to be noted that the following integral needs to be solved

\begin{equation}\label{b33}
I=\int \psi_{m_1,m_2}^{\ast(0)}(q_1,p_1,q_2,p_2)\frac{B^2}{8}[(\widehat{a}+\widehat{a}^{\dagger})^2+(\widehat{b}+\widehat{b}^{\dagger})^2]^3\psi_{n_1,n_2}^{(0)}(q_1,p_1,q_2,p_2)dq_1dp_1dq_2dp_2\,,
\end{equation}
before a solution for Eq.(\ref{b32}). Using the orthogonality relations 
\begin{equation}\label{ortob1}
\int \phi_{n}^{\ast}(q_1,p_1)\phi_{m}(q_1,p_1)dq_1dp_1=\delta_{n,m},
\end{equation}
\begin{equation}\label{ortob2}
\int \Gamma_{n}^{\ast}(q_2,p_2)\Gamma_{m}(q_2,p_2)dq_2dp_2=\delta_{n,m},
\end{equation}
the ground-state is

\begin{eqnarray}\label{b40}
\psi^{(1)}_{0,0}(q_1,p_1,q_2,p_2)&=&\psi^{(0)}_{0,0}(q_1,p_1,q_2,p_2)\\\nonumber&+&\frac{B^2}{8W}\Big[\left(-21\sqrt{2}-18-25\sqrt{10}\right)\psi^{(0)}_{2,0}\\\nonumber&+&\left(-3\frac{\sqrt{2}}{2}-3\right)\psi^{(0)}_{2,2}+(-30\sqrt{21}-3\sqrt{6})\psi^{(0)}_{4,0}\\\nonumber&+&4\sqrt{3}\psi^{(0)}_{4,2}-8\sqrt{1155}\psi^{(0)}_{6,0}\Big]\nonumber.
\end{eqnarray}
And for excited states, the wave functions are 
\begin{eqnarray}\label{b41a}
\psi^{(1)}_{1,0}&=&\psi^{(0)}_{1,0}+\frac{B^2}{8W}\Big[89,30\psi^{(0)}_{1,2}+13,47\psi^{(0)}_{1,4}+6,32\psi^{(0)}_{1,6}-89,43\psi^{(0)}_{3,0}\\\nonumber
&-&19,33\psi^{(0)}_{3,2}-23,51\psi^{(0)}_{5,0}-10,31\psi^{(0)}_{5,4}-11,83\psi^{(0)}_{7,0}\Big],\nonumber
\end{eqnarray}
and
\begin{eqnarray}\label{b41b}
\psi^{(1)}_{0,1}&=&\psi^{(0)}_{0,1}+\frac{B^2}{8W}\Big[-89,30\psi^{(0)}_{2,1}-13,47\psi^{(0)}_{4,1}-6,32\psi^{(0)}_{6,1}+89,43\psi^{(0)}_{0,3}\\\nonumber
&+&19,33\psi^{(0)}_{2,3}+23,51\psi^{(0)}_{0,5}+10,31\psi^{(0)}_{4,5}+11,83\psi^{(0)}_{0,7}\Big].\nonumber
\end{eqnarray}

The Wigner function for the hydrogen atom in a constant magnetic field is given by
\begin{equation}\label{wigb}
f_W(q_1,p_1,q_2,p_2)=\psi_{n_1,n_2}^{\ast(1)}(q_1,p_1,q_2,p_2)\star \psi_{n_1,n_2}^{(1)}(q_1,p_1,q_2,p_2).
\end{equation}

\begin{figure}[!htb]
\centering
\includegraphics[scale=0.6]{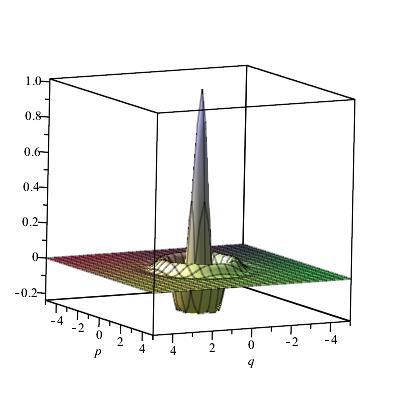}
\caption{Wigner function zero order - Zeeman effect, $E=1$ and $B=1$}
\label{z1}
\end{figure}

\begin{figure}[!htb]
\centering
\includegraphics[scale=0.6]{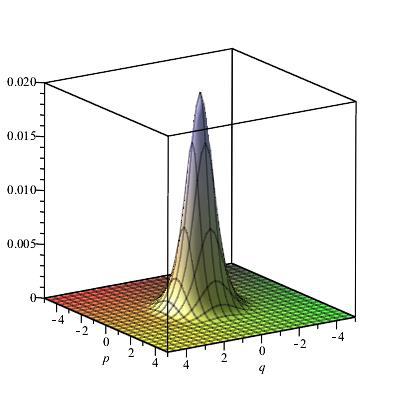}
\caption{Wigner function zero order - Zeeman effect, $E=1$ and $B=0.1$}
\label{z2}
\end{figure}

\begin{figure}[!htb]
\centering
\includegraphics[scale=0.6]{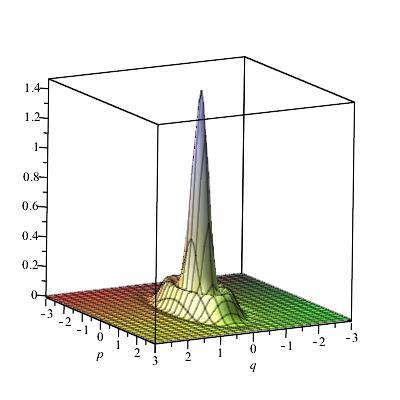}
\caption{Wigner function first order - Zeeman effect, $E=10$ and $B=1$}
\label{z3}
\end{figure}

\begin{figure}[!htb]
\centering
\includegraphics[scale=0.6]{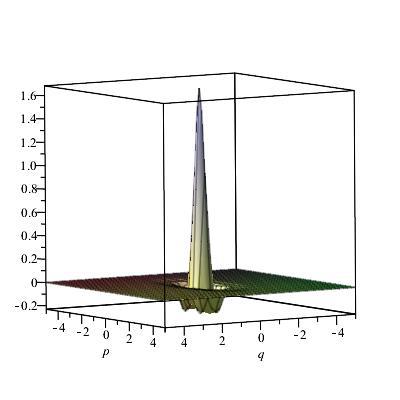}
\caption{Wigner function first order - Zeeman effect, $E=10$ and $B=0.1$}
\label{z4}
\end{figure}

It should be noted that all plots consider $q_2=p_2=1$ in order to show a 3D figure, thus $q_1=q$ and $p_1=p$. In Fig.(\ref{z1}) and Fig.(\ref{z2}), the behavior of Wigner function is presented for order zero with magnetic field assuming values $B=1$ and $B=0.1$, respectively with $E=1$. In Fig.(\ref{z3}) and Fig.(\ref{z4}), the behavior of Wigner function is shown for first order with magnetic field assuming values $B=1$ and $B=0.1$, respectively, with $E=10$.

Comparing the graphics given in Figs. (\ref{z1})-(\ref{z4}), the negative part of the Wigner function increases with larger values of energy and magnetic field.

\begin{figure}[!htb]
\centering
\includegraphics[scale=0.6]{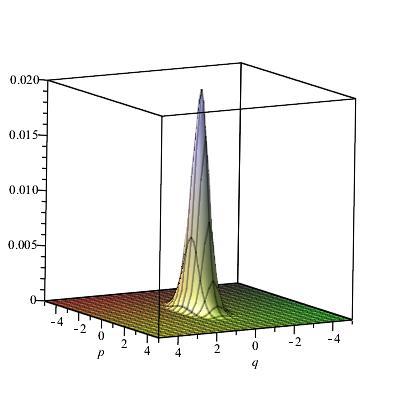}
\caption{Wigner function first order - Zeeman effect, $E=1$ and $B=1$}
\label{z5}
\end{figure}

\begin{figure}[!htb]
\centering
\includegraphics[scale=0.6]{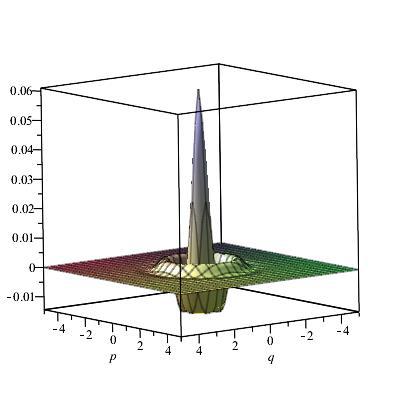}
\caption{Wigner function first order - Zeeman effect, $E=1$ and $B=0.1$}
\label{z6}
\end{figure}

\begin{figure}[!htb]
\centering
\includegraphics[scale=0.6]{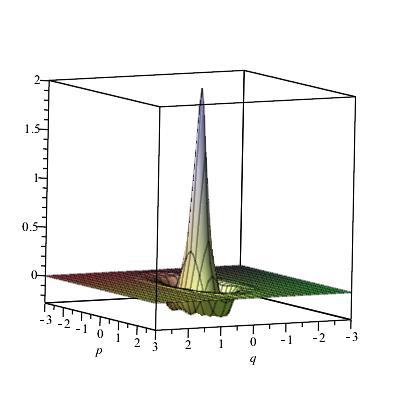}
\caption{Wigner function first order - Zeeman effect, $E=1$ and $B=0.5$}
\label{z7}
\end{figure}

\begin{figure}[!htb]
\centering
\includegraphics[scale=0.6]{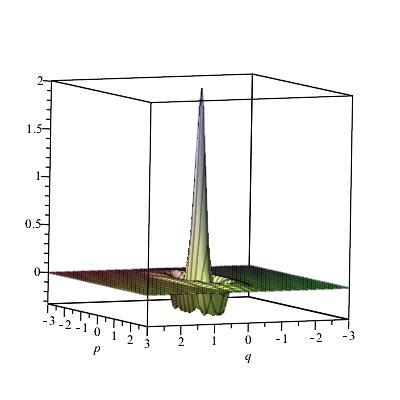}
\caption{Wigner function first order - Zeeman effect, $E=10$ and $B=0.5$}
\label{z8}
\end{figure}

Wigner function to first order for magnetic field values $B=1$ and $B=0.1$ is shown in Figs. (\ref{z5}) and Fig.(\ref{z6}) for $E=1$. The behavior of Wigner function to first order  with magnetic field value $B=0.5$ is shown in Figs. (\ref{z7}) and Fig.(\ref{z8}), for $E=1$ and $E=10$, respectively.

 The correction of first order of engeivalue of Eq.(\ref{b23}) is given by
\begin{equation}\label{energia1}
\delta k^{(1)}_{n_1,n_2}=\int \psi_{n_1,n_2}^{(0)}\widehat{H}_1\psi_{n_1,n_2}^{\ast(0)}dq_1dp_1dq_2dp_2.
\end{equation}

Performing calculations for $\widehat{H}_1$ leads to

\begin{equation}\label{enecor}
k^{(1)}_{n_1,n_2}=(n_1+n_2+1)W+\frac{B^2}{8}\delta,
\end{equation}
where
\begin{eqnarray}\label{delta}
\delta&=&(n_1+1)(n_1+2)(n_1+3)\\\nonumber
&+&(n_1+1)(n_1+2)^2+(n_1-1)n_1(n_1+1)^2\\\nonumber
&+&(n_1+1)n_1(n_1+1)+\sqrt{n_1^3(n_1+1)^3}\\\nonumber
&+&(n_1+1)n_1^2+(n_1+1)n_1(n_1-1)\\\nonumber
&+&n_1(n_1-1)^2+n_1(n_1-1)(n_1-2)\\\nonumber
&+&3(n_1+1)n_1(n_2+1)+3(n_1+1)n_1n_2\\\nonumber
&+&3n_1^2(n_2+1)+3n_1^2n_2+3(n_1-1)n_1(n_2+1)\\\nonumber
&+&3(n_1-1)n_1n_2+3(n_2+1)n_2(n_1+1)\\\nonumber
&+&3(n_2+1)n_2n_1+3n_2^2(n_1+1)+3n_2^2n_1\\\nonumber
&+&3(n_2-1)n_2(n_1+1)+3(n_2-1)n_2n_1\\\nonumber
&+&(n_2+1)(n_2+2)(n_2+3)+(n_2+1)(n_2+2)^2\\\nonumber
&+&(n_2-1)n_2(n_2+1)^2+\sqrt{n_2^3(n_2+1)^3}\\\nonumber
&+&\sqrt{(n_2+1)^2n_2^4}+(n_2+1)n_2(n_2-1)\\\nonumber
&+&n_2(n_2-1)^2 + n_2(n_2-1)(n_2-2).\nonumber
\end{eqnarray}
Then the eigenvalue is 
\begin{equation}\label{valorw}
W=\frac{k^{(1)}_{n_1,n_2}-(B^2/8)\delta}{n_1+n_2+1}.
\end{equation}

With the relation $W^2/2=-E$, the eigenvalues associated to the Zeeman effect in phase space are given by

\begin{equation}\label{enefim}
E_{n_1,n_2}=-\frac{1}{2}\left[\frac{k^{(1)}_{n_1,n_2}-(B^2/8)\delta}{n_1+n_2+1}\right]^2\,,
\end{equation}
and
\begin{equation}\label{enefimb}
E_{N}=-\frac{1}{2}\left[\frac{k^{(1)}_{n_1,n_2}-(B^2/8)\delta}{N}\right]^2\,,
\end{equation}
where $N=n_1+n_2+1$. Note that if $B\rightarrow 0$, the known results are obtained \cite{campos}.

Using the Wigner function, the negative parameter for the system is calculated. The results are presented in Tables (\ref{tbl:tablelabelnc2}-\ref{tbl:tablelabelnc3}).
It is to be noted that when the magnetic field increases the negativity parameter also increases. In addition, for a given value of the magnetic field, the negativity parameter increases when the sum $n_1+n_2$ increases. This result is presented in the graphics above.

\begin{table}
\caption{Negativity parameter, $B=1$.}
\centering
\begin{tabular}{ll}
\hline
$n_1,n_2$ & $\eta(\psi)$ \\
\hline
$0,0$  & $0.14345$\\
$0,1$ & $0.32645$\\
$1,0$ & $0.32645$\\
$1,1$ & $0.45786$\\
$2,0$ & $0.45786$\\
$0,2$ & $0.45786$\\
\hline
\end{tabular}
\label{tbl:tablelabelnc2}
\end{table}

\begin{table}
\caption{Negativity parameter, $B=0.1$.}
\centering
\begin{tabular}{ll}
\hline
$n_1,n_2$ & $\eta(\psi)$ \\
\hline
$0,0$  & $0.0034$\\
$1,0$ & $0.0562$\\
$0,1$ & $0.0562$\\
$1,1$ & $0.0635$\\
$2,0$ & $0.0635$\\
$0,2$ & $0.0635$\\
\hline
\end{tabular}
\label{tbl:tablelabelnc3}
\end{table}

\section{Concluding remarks}\label{sec4}
The Zeeman effect in phase space for Schr\"odinger equation, which endows the Galilean symmetry, is analyzed. The Wigner function is calculated numerically and presented in the panels for several parameters. Such a function has a clear interpretation in the classical limit and can be projected in the momenta or coordinate space for experimental purpose. The modulus of the Wigner function is also finite that allows a calculation its negativity. The results are presented in Tables (\ref{tbl:tablelabelnc2}-\ref{tbl:tablelabelnc3}). It indicates a direct relation between the magnetic field and the discrete parameter $N=n_1+n_2$. The increase of the magnetic field is related to the departure from the classical behavior since the negativity parameter increases accordingly.

\section*{Acknowledgements}
This work was partially supported by CNPq of Brazil.

\section*{Data Availability}
No experimental data were used in this article.

\end{document}